\documentclass[aps,pra,reprint,superscriptaddress,showkeys,longbibliography]{revtex4-2} % {{{*
\usepackage[T1]{fontenc}
\usepackage{amsmath,amssymb}
\usepackage{graphicx}
\usepackage{physics}
\usepackage{microtype}
\usepackage[nolinenumbers,noshowlabels]{jadeleon} % {{{*
\DeclareCommenter{mg}{0057B8}{MG}
\DeclareCommenter{pv}{008000}{PV}
\usepackage{hyperref}
\usepackage{orcidlink}
\hypersetup{colorlinks=true,linkcolor=blue,citecolor=blue,urlcolor=blue}
\newcommand{\cT}{\mathcal T}

\newcommand{\icn}{Instituto de Ciencias Nucleares, \unam}
\begin{document}

\title{Reachability Constraints and Monotones for Quantum-State Texture} % {{{*
% Add author names, affiliations, and corresponding-author details here.
\author{Jose Alfredo de Leon\,\orcidlink{0000-0003-0045-9017}}
\email{deleongarrido.jose@gmail.com}
\affiliation{\ifunam}
\author{Pablo Vielman}
\affiliation{Escuela de Ciencias Físicas y Matemáticas, Universidad de San Carlos de Guatemala, Ciudad Universitaria, Guatemala 01012, Guatemala}
\author{Miguel Gonzalez\,\orcidlink{0009-0004-0112-5988}}
\email{miguel.gonzalez@correo.nucleares.unam.mx}
\affiliation{\icn}
%\affiliation{Okinawa Institute of Science and Technology Graduate University (OIST), Okinawa, Japan}
% *}}}
\begin{abstract} % {{{*
How a quantum resource can be quantified depends on which operations are free.
Yet two proposed quantifiers of quantum state texture can increase
under free operations: the Wigner--Yanase--Dyson skew information
relative to the unique textureless state and the squared Hilbert--Schmidt
distance from that state.
We show reachability from every state orthogonal to the
textureless state, whether pure or mixed: each can be freely transformed
into any other state.
Therefore, every texture monotone must assign the same maximal value to
all such states, a requirement that neither candidate satisfies.
We then construct valid texture monotones by adding the population
outside the textureless state to the skew information and by maximizing
the squared Hilbert--Schmidt distance over free operations.
We prove faithfulness, convexity, and invariance under textureless
ancillas, characterize additivity properties and strong monotonicity, satisfied
only in the Wigner--Yanase case, and derive tight bounds for both
monotones at fixed overlap with the textureless state and characterize
all saturating states.
Together, these results show how allowed state transformations impose 
operational constraints on the quantification of quantum state texture.
\end{abstract}% *}}}
% \keywords{Quantum state texture, Wigner--Yanase--Dyson skew information, Resource theory}
\maketitle

\section{Introduction} % {{{*

Resource theories relate restrictions on state manipulation to the
capabilities of quantum states in information-processing
tasks~\cite{ChitambarGour2019}. For a closed convex set of free states,
every resource state can outperform all free inputs in a suitably
chosen channel-discrimination task~\cite{Takagi2019}. Beyond convexity,
families of multicopy channel-discrimination tasks also establish
operational advantages for every resource state~\cite{Kuroiwa2024}.
% These connections motivate quantifying texture with measures that
% respect the allowed state transformations. 
These operational advantages has motivated identifying which features of quantum 
states can be treated as resources.

Quantum state texture (QST) is a quantum resource theory in which a quantum
state's departure from a flat state in a chosen basis is elevated to a
resource~\cite{Parisio2024}. QST is specified
by its free states, free operations, and resource monotones.
Its unique free state is the \emph{textureless state} $f_1$, which is the
uniform superposition in a reference basis.
The free operations, denoted by $\Lambda$, are completely positive,
trace-preserving maps that have $f_1$ as a fixed point,
$\Lambda(f_1)=f_1$. A texture monotone $\cT$
quantifies the resource: it must be nonnegative, vanish at $f_1$, and
never increase under free operations~\cite{ChitambarGour2019}.
% Faithfulness additionally requires that $\cT(\rho)=0$ only if $\rho=f_1$.
% Convexity is a desirable additional property, adopted as an axiom in the
% original QST formulation~\cite{Parisio2024}.

Several texture quantifiers have been proposed to quantify quantum state texture.
The original rugosity is a function of the overlap with the textureless
state~\cite{Parisio2024}; this overlap has been measured for qubits
in nuclear magnetic resonance experiments~\cite{Vieira2026}.
Subsequent proposals use distance, geometry, fidelity, relative entropy,
and robustness~\cite{Wang2025}; weight and Tsallis relative
entropy~\cite{Zhang2025}; generalized relative
entropies~\cite{Cao2026,ChenLei2026}; affinity, Tsallis relative entropy,
and convex-roof constructions~\cite{Cui2025}; and Hellinger distance
and Jensen--Shannon divergence~\cite{Muthuganesan2026}.
Convex-function-based quantifiers and texture cost have also been introduced~\cite{Lin2026},
while a grand-sum-based monotone and set texture have also been
studied~\cite{deLeon2026}.
Related extensions include texture-based quantifiers of other resources
and nonlocal texture~\cite{Patra2026}, generalized reference states in
fixed-point resource theories~\cite{Greenwood2026}, and quantifiers of
quantum-state block texture~\cite{Chu2026Block}.

Reachability under free operations constrains all these proposals:
if a free operation transforms $\rho$ into $\sigma$, every texture
monotone must satisfy $\cT(\rho)\ge\cT(\sigma)$~\cite{ChitambarGour2019}.
Consequently, if every state in a set can reach any target state by a
free operation, every texture monotone must assign the same global
maximum to all states in that set.
In QST, we build upon results of Refs.~\cite{Parisio2024,Ma2026Hawking}
to show this set contains every state supported on the orthogonal
complement of the textureless state. We call these states \emph{maximally textured
states}; they include pure states and, in dimensions greater than two, mixed states.
Their common maximal value is required for any texture
monotone, providing a necessary
condition for assessing proposed quantifiers.

Using this reachability from the maximally textured states, we show why two proposed texture
quantifiers in the literature fail monotonicity and construct valid alternatives from the
same quantities.
The Wigner--Yanase--Dyson (WYD) skew information relative to the textureless
state~\cite{Muthuganesan2026} vanishes on every maximally textured state
but is positive on some other states, so it cannot assign the former
the global maximum. The squared Hilbert--Schmidt distance from the
textureless state~\cite{Cui2025} assigns different values to freely
interconvertible maximally textured states in dimensions greater than
two, violating the requirement of a common maximum.
For the WYD construction, adding the population outside the textureless
state yields a family of texture monotones. For the Hilbert--Schmidt
construction, maximizing the squared distance over free operations
yields a texture monotone with a closed-form expression.
Both corrected constructions are faithful, convex, and monotone under
free operations.

The paper is organized as follows. After setting out the QST framework
and the requirements for texture measures in \Sref{sec:preliminaries}, we
establish reachability from maximally textured states in
\Sref{sec:reachability}. In \Sref{sec:proposed-measures}, we use these
reachability constraints to show why the proposed WYD and squared Hilbert--Schmidt
quantities fail monotonicity. \Sref{sec:wyd-monotones} and
\Sref{sec:hs-monotone} introduce the corrected measures, establish their
faithfulness, convexity, and monotonicity, and examine strong monotonicity
and behavior under tensor products. \Sref{sec:texture-bounds}
uses the free twirl and its isotropic axis to motivate bounds at fixed
overlap with the textureless state, determines the exact ranges of both
corrected measures, characterizes the saturating states, and shows that
all intermediate values are attainable. We summarize the results in
\Sref{sec:conclusions}. Appendices~\ref{app:hs-optimization}
and~\ref{app:twirl} provide the optimal free channel for the
Hilbert--Schmidt measure and the properties of the free twirl,
respectively. Appendix~\ref{app:wyd-strong-counterexample} gives the
three-dimensional counterexample to strong monotonicity for the WYD family.
% *}}}
\section{Preliminaries} % {{{*
\label{sec:preliminaries}

Quantum state texture is defined relative to a fixed orthonormal basis
$\{\ket j\}_{j=0}^{D-1}$ of a finite-dimensional Hilbert space, with $D\ge2$.
States are represented by density operators $\rho,\sigma\succeq0$ of unit trace (we write $\sigma\succeq0$ for a positive semidefinite operator and $\sigma>0$ for a positive definite operator).
The unique textureless state is the uniform superposition in this
basis~\cite{Parisio2024},
\begin{equation}
 \ket{f_1}=\frac1{\sqrt D}\sum_{j=0}^{D-1}\ket j,
 \qquad f_1=\dyad{f_1}.
 \label{eq:prelim-textureless}
\end{equation}
The overlap $p=\Tr(\rho f_1)=\mel{f_1}{\rho}{f_1}$ is the population
in the textureless subspace. By \Eref{eq:prelim-textureless}, it satisfies
$p=\Sigma(\rho)/D$, where $\Sigma(\rho)=\sum_{j,k}\mel{j}{\rho}{k}$ is
the grand sum of the density matrix in the reference basis~\cite{deLeon2026}. 
The complementary projector $\mathbb I-f_1$ selects the orthogonal
subspace, whose normalized vectors will be denoted by $\ket g$.

Free operations are quantum channels that preserve the textureless state.
A quantum channel $\mathcal E$ is a completely positive, trace-preserving
(CPTP) map with a Kraus representation~\cite{Watrous2018}
\begin{equation}
 \mathcal E(X)=\sum_jK_jXK_j^\dagger,
 \qquad \sum_jK_j^\dagger K_j=\mathbb I.
 \label{eq:prelim-kraus}
\end{equation}
We denote QST free channels by $\Lambda$; their defining condition is
$\Lambda(f_1)=f_1$~\cite{Parisio2024}. For a specified Kraus realization, outcome $j$ occurs
with probability $p_j=\Tr(K_j\rho K_j^\dagger)$ and leaves the conditional
state $\rho_j=K_j\rho K_j^\dagger/p_j$ when $p_j>0$.
Discarding the result gives $\mathcal E(\rho)=\sum_jp_j\rho_j$;
retaining it gives access to the conditional states. 
Channels that measure the input and prepare an output state for each
result~\cite{HorodeckiShorRuskai2003} will be used to establish
reachability under free operations.

A texture monotone must respect the state transformations allowed by
these free operations. A texture monotone $\cT$ must be
nonnegative on all states, vanish at the textureless state, and never
increase under a free operation $\Lambda$~\cite{ChitambarGour2019}:
\begin{equation}
 \cT(\rho)\ge0,\qquad \cT(f_1)=0,\qquad
 \cT(\Lambda(\rho))\le\cT(\rho).
 \label{eq:prelim-monotone}
\end{equation}
These requirements allow states to be compared by their resource content. 
An overlap-based benchmark used below is the geometric
texture $\cT_g(\rho)=1-\Tr(\rho f_1)$~\cite{Wang2025}.

We distinguish these basic requirements from additional desirable properties of a
texture monotone~\cite{ChitambarGour2019}. Faithfulness means that $\cT(\rho)=0$ only for $\rho=f_1$, so every
textured state is assigned a positive value.
Convexity and strong monotonicity require, respectively,
\begin{equation}
 \begin{aligned}
 \cT\Bigl(\sum_jp_j\rho_j\Bigr)&\le\sum_jp_j\cT(\rho_j),\\
 \sum_jp_j\cT(\rho_j)&\le\cT(\rho).
 \end{aligned}
 \label{eq:prelim-additional-properties}
\end{equation}
The first inequality concerns any ensemble with $p_j\ge0$ and
$\sum_jp_j=1$; in the second, $p_j$ and $\rho_j$ refer to the outcomes
of a free operation.
Convexity expresses that forgetting which state was prepared cannot
increase texture above the preparation average. Strong monotonicity
requires that resolving the outcomes of a free operation cannot increase
texture on average, although an individual selected outcome may have
more texture than the input.

For composite systems, we use the product textureless state
$f_1=f_1^{(A)}\otimes f_1^{(B)}$, where
$f_1^{(X)}=\dyad{f_1^{(X)}}$, and denote the partial trace over $B$ by
$\Tr_B$. Additivity means
$\cT(\rho\otimes\sigma)=\cT(\rho)+\cT(\sigma)$;
subadditivity and superadditivity replace this equality by $\le$ and
$\ge$, respectively. Invariance under adding a textureless ancilla means
$\cT(\rho_A\otimes f_1^{(B)})=\cT(\rho_A)$.

The construction in \Sref{sec:wyd-monotones} uses the
Wigner--Yanase--Dyson (WYD) skew information~\cite{Hansen2008},
which for an observable $H$ reads
\begin{equation}
 I_\alpha(\rho,H)=\Tr(\rho H^2)
 -\Tr(\rho^\alpha H\rho^{1-\alpha}H),\qquad 0<\alpha<1.
 \label{eq:prelim-wyd}
\end{equation}
Fractional powers are defined spectrally: if
$\rho=\sum_j\lambda_j\dyad{e_j}$, then
$\rho^s=\sum_j\lambda_j^s\dyad{e_j}$ for $s>0$.
The case $\alpha=1/2$ is the Wigner--Yanase skew information.
For pure states, $I_\alpha$ equals the variance of $H$; it vanishes when
$\rho$ and $H$ commute.
A proposed texture quantity takes $H=f_1$~\cite{Muthuganesan2026}, so it probes quantum
uncertainty relative to the textureless projector.

The Hilbert--Schmidt construction in \Sref{sec:hs-monotone} uses distances
from the textureless state.
The Hilbert--Schmidt and trace norms are
$\norm{X}_2^2=\Tr(X^\dagger X)$ and
$\norm{X}_1=\Tr\sqrt{X^\dagger X}$, respectively.
They define the squared Hilbert--Schmidt distance
$\norm{\rho-\sigma}_2^2$ and the trace distance
$D_{\mathrm{tr}}(\rho,\sigma)=\tfrac12\norm{\rho-\sigma}_1$.
The former depends on the purities of the two states and their overlap;
in particular, its value relative to $f_1$ contains the purity
$\Tr(\rho^2)$. The trace distance is contractive under every CPTP map~\cite{Watrous2018},
whereas the Hilbert--Schmidt distance has no such general
property~\cite{Ozawa2000}.
To assess these candidate texture quantities, we first establish which
state transformations are possible under free operations. 
% *}}}
\section{Reachability from states orthogonal to the textureless state} % {{{*
\label{sec:reachability}
The Fourier states orthogonal to the textureless state were identified
as maximally resourceful states, from which any target state can be
reached by free operations~\cite{Parisio2024}. Every pure state
orthogonal to $\ket{f_1}$ is related to one of these Fourier states by
a free unitary and therefore has the same maximal texture~\cite{Lin2026}.
It was subsequently recognized that this maximality also holds for every
state supported on the orthogonal complement of $\ket{f_1}$, including mixed states
supported on that subspace~\cite{Ma2026Hawking}.

The measure-and-prepare construction for free interconvertibility in the
orthogonal complement also allows arbitrary target states.
Let us denote this set by
$\mathcal S_\perp=\{\rho:\Tr(\rho f_1)=0\}$. For a system with
$D$-dimensional Hilbert space, the orthogonal complement of $\ket{f_1}$
has dimension $D-1$: it is one-dimensional for a qubit and has higher
dimension for $D>2$. This interconvertibility within $\mathcal S_\perp$ can be established by a
measure-and-prepare channel whose prepared target also belongs to
this subspace~\cite{Ma2026Hawking}. The reachability beyond
this set is achieved simply by allowing the prepared target $\sigma$ to be any density
operator $\sigma$ on the full Hilbert space:
\begin{equation}
  \Lambda_\sigma(\rho)
  =\Tr(\rho f_1)f_1 + \Tr[\rho(\mathbb I-f_1)]\sigma,
  \label{eq:measure-prepare-free}
\end{equation}
where $\mathbb I$ is the identity operator on the system's Hilbert space.
This map is CPTP because it is a measure-and-prepare
channel~\cite{HorodeckiShorRuskai2003}. It performs the projective
measurement $\{f_1,\mathbb I-f_1\}$, which distinguishes the
textureless state from its orthogonal complement, and prepares $f_1$
or $\sigma$, respectively. It fixes $f_1$ and is therefore a QST free operation.

We call states supported on the orthogonal complement of $\ket{f_1}$
\emph{maximally textured states}, since each can be transformed into any
target state by a free operation.
For any $\rho_\perp\in\mathcal S_\perp$, the first measurement
outcome in \Eref{eq:measure-prepare-free} has probability zero and the second has
probability one. Consequently, $\Lambda_\sigma(\rho_\perp)=\sigma$.
Since both $\rho_\perp$ and $\sigma$ are arbitrary, every state is
reachable from every state in $\mathcal S_\perp$ by a free operation.
The Fourier states orthogonal to $\ket{f_1}$ are only a subset of
$\mathcal S_\perp$: this set contains every pure state in the orthogonal
subspace and, for $D>2$, mixed states supported on it as well
[\Eref{eq:measure-prepare-free}].

Every texture monotone must assign the same global maximum to all
maximally textured states. The QST free map in \Eref{eq:measure-prepare-free}
transforms any $\rho_\perp\in\mathcal S_\perp$ into any target
state $\sigma$, so monotonicity requires
\begin{equation}
 \cT(\sigma)=\cT(\Lambda_\sigma(\rho_\perp))
 \le\cT(\rho_\perp)
 \label{eq:orthogonal-maximality}
\end{equation}
for every target $\sigma$ [\Eref{eq:prelim-monotone}]. In particular,
transformations in both directions between states in this set force
their values to be equal.

The common maximum required by \Eref{eq:orthogonal-maximality} provides
a criterion for testing proposed texture quantifiers, which we apply in
\Sref{sec:proposed-measures}.

% *}}}
\section{Failure of two proposed texture quantifiers} % {{{*
\label{sec:proposed-measures}

A nonnegative monotone that vanishes at even one maximally textured state
must vanish identically. Indeed, if $\cT(\rho_\perp)=0$,
\Eref{eq:orthogonal-maximality} and nonnegativity force
$\cT(\sigma)=0$ for every state $\sigma$. We now show that a proposed WYD quantity and a proposed squared Hilbert--Schmidt
quantity violate, respectively, this
condition on their zeros and the requirement of a common maximal value
on $\mathcal S_\perp$.

\subsection{WYD skew information} % {{{*

A proposed WYD quantity vanishes on all states that commute with the
textureless state, including every maximally textured state. It is defined as~\cite{Muthuganesan2026}
\begin{equation}
 \begin{aligned}
 \cT^{\mathrm{WYD}}_\alpha(\rho)
 &:=I_\alpha(\rho,f_1)\\
 &=\Tr(\rho f_1)
   -\Tr(\rho^\alpha f_1\rho^{1-\alpha}f_1),
 \end{aligned}
 \label{eq:wyd}
\end{equation}
where $0<\alpha<1$ and $I_\alpha(\rho,f_1)$ is the WYD skew information
of $\rho$ with respect to the projector $f_1$. It quantifies quantum
uncertainty in the projective measurement $\{f_1,\mathbb I-f_1\}$,
whose variance is $\Tr(\rho f_1)-[\Tr(\rho f_1)]^2$. For pure states,
the skew information equals this variance; for mixed states, it is
bounded above by it. Whenever $\comm{\rho}{f_1}=0$, the two trace
terms coincide, giving $\cT^{\mathrm{WYD}}_\alpha(\rho)=0$ for every
$0<\alpha<1$. Since every state in $\mathcal S_\perp$ commutes with
$f_1$, this quantity vanishes on every maximally textured state.

\begin{figure}[t]
 \centering
  \includegraphics[width=\columnwidth]{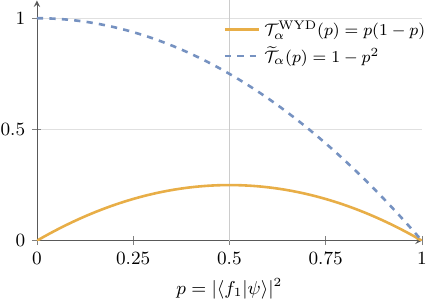}
 \caption{The function $\cT^{\mathrm{WYD}}_\alpha$,
 \Eref{eq:wyd}, and the WYD-ske-information--based monotone $\widetilde\cT_\alpha$,
 \Eref{eq:corrected}, for pure states $\ket{\psi}$ as functions of the overlap
 with the textureless state, $p=\abs{\braket{f_1}{\psi}}^2$, for every $0<\alpha<1$ and $D\ge2$.
 They reduce to $p(1-p)$ (solid) and $1-p^2$ (dashed), respectively.
 The function $\cT^{\mathrm{WYD}}_\alpha$ vanishes at the textureless state and at states
 orthogonal to it, but is positive for $0<p<1$, hence failing to be a texture monotone (see main text).}
 \label{fig:wyd-pure-states}
\end{figure}

The reachability criterion rules out the quantity in \Eref{eq:wyd} as a texture monotone
because it is not identically zero. Indeed, \Eref{eq:orthogonal-maximality}
is incompatible with its zeros on $\mathcal S_\perp$. For a normalized
$\ket{g}\perp\ket{f_1}$ and
$\ket{\psi}=(\ket{f_1}+\ket{g})/\sqrt2$, one has
$I_\alpha(\dyad{\psi},f_1)=1/4$.
The channel in \Eref{eq:measure-prepare-free}, with
$\sigma=\dyad{\psi}$, therefore gives the explicit violation
\begin{equation}
 \begin{aligned}
 I_\alpha(\Lambda_\sigma(\rho_\perp),f_1)
 &=\tfrac14\\
 &>0=I_\alpha(\rho_\perp,f_1)
 \end{aligned}
 \label{eq:counterexample}
\end{equation}
for any $\rho_\perp\in\mathcal S_\perp$. Thus a proposed quantity
fails monotonicity for every $0<\alpha<1$ and every dimension $D\ge2$.
This recovers the known refutation and makes explicit that the qubit
counterexample remains valid in arbitrary dimension~\cite{Lin2026}.
For pure states, \fref{fig:wyd-pure-states} illustrates this failure:
the quantity equals $p(1-p)$, where $p=\Tr(\rho f_1)$, vanishing at
zero and unit overlap but remaining positive in between.
The reachability criterion also excludes any other nonnegative,
nontrivial candidate that vanishes on even one maximally textured state.
In particular, detecting noncommutativity with $f_1$ alone cannot define
a texture monotone.

% *}}}
\subsection{Squared Hilbert–Schmidt distance} % {{{*

The squared Hilbert--Schmidt distance proposed in Ref.~\cite{Cui2025}
also fails to be a texture monotone in dimensions greater than two.
It is a texture monotone only for qubits, as established below in
\Eref{eq:hs-qubit-trace-distance}. In our notation,
this candidate is
\begin{equation}
 \begin{aligned}
 \cT_{\mathrm{HS}}(\rho)
 &:=\norm{\rho-f_1}_2^2\\
 &=1+\Tr(\rho^2)-2\Tr(\rho f_1),
 \end{aligned}
 \label{eq:hs-texture}
\end{equation}
where $\norm{A}_2$ is the Hilbert--Schmidt norm, defined by
$\norm{A}_2^2=\Tr(A^\dagger A)$. This quantity is nonnegative,
faithful, and convex, but these properties do not ensure monotonicity.
For any maximally textured state $\rho_\perp$, it reduces to
$\cT_{\mathrm{HS}}(\rho_\perp)=1+\Tr(\rho_\perp^2)$.
For $D\ge3$, the purity varies within $\mathcal S_\perp$, whereas
\Eref{eq:orthogonal-maximality} requires every texture monotone to
have the same value throughout this set.

A QST free operation that purifies a maximally textured state gives an
explicit violation. For $D\ge3$, let us choose a normalized $\ket{g}\perp\ket{f_1}$ and
let us set
\begin{equation}
 \rho_\perp=\frac{\mathbb I-f_1}{D-1},
 \qquad \sigma=\dyad{g}.
 \label{eq:hs-counterexample-states}
\end{equation}
The channel $\Lambda_\sigma$ in \Eref{eq:measure-prepare-free} fixes
$f_1$ and maps $\rho_\perp$ to $\sigma$. Both states have zero overlap
with $f_1$, while their purities are $1/(D-1)$ and $1$, respectively.
\Eref{eq:hs-texture} therefore gives
\begin{equation}
 \begin{aligned}
 \cT_{\mathrm{HS}}(\Lambda_\sigma(\rho_\perp))
 =2
 >1+\frac{1}{D-1}=\cT_{\mathrm{HS}}(\rho_\perp).
 \end{aligned}
 \label{eq:hs-monotonicity-violation}
\end{equation}
For a qutrit, the increase is from $3/2$ to $2$. The purity term in
\Eref{eq:hs-texture} thus assigns different values to states that
must all attain the same global maximum by
\Eref{eq:orthogonal-maximality}.

The proposed justification of monotonicity does not account for this
increase in purity. The argument in Ref.~\cite{Cui2025} assumes
$\norm{\Lambda(\rho-f_1)}_2=\norm{\rho-f_1}_2$ for QST free operations.
The fixed-point condition $\Lambda(f_1)=f_1$ does not imply this equality
or even the inequality required for monotonicity.

The qubit case is an exception: the squared Hilbert--Schmidt distance to
$f_1$ is monotone under every free qubit operation. Indeed, any traceless
Hermitian $2\times2$ matrix has eigenvalues $a$ and $-a$, so
\begin{equation}
 \norm{\rho-f_1}_2^2=\frac12\norm{\rho-f_1}_1^2,
 \label{eq:hs-qubit-trace-distance}
\end{equation}
where $\norm{A}_1=\Tr\sqrt{A^\dagger A}$ is the trace norm. Contractivity of the trace
norm under CPTP maps, together with $\Lambda(f_1)=f_1$, then proves
monotonicity. This qubit property does not extend to higher dimensions,
where \Eref{eq:hs-monotonicity-violation} rules out a proposed
quantity as a texture monotone.

% *}}}
% *}}}
\section{A family of WYD texture monotones} % {{{*
\label{sec:wyd-monotones}
For $0<\alpha<1$, adding the population outside the textureless state
turns a proposed WYD quantity into a texture monotone. The WYD texture family is
\begin{equation}
 \begin{aligned}
 \widetilde\cT_\alpha(\rho)
 &:=I_\alpha(\rho,f_1)+1-\Tr(\rho f_1)\\
 &=1-q_\alpha(\rho)q_{1-\alpha}(\rho),\\
 q_\alpha(\rho)&:=\mel{f_1}{\rho^\alpha}{f_1},\qquad 0<\alpha<1,
 \end{aligned}
 \label{eq:corrected}
\end{equation}
The added population term, $1-\Tr(\rho f_1)$, assigns unit texture to
every maximally textured state, as reachability requires.
For a pure state with overlap $p=\Tr(\rho f_1)$, the family reduces to
$1-p^2$, independently of $\alpha$. For mixed states, the two spectral
moments $q_\alpha$ and $q_{1-\alpha}$ also retain information about the
spectrum of $\rho$ and its orientation relative to $\ket{f_1}$.

The WYD texture family satisfies the three basic requirements of a texture
monotone: nonnegativity, vanishing at the textureless state, and
monotonicity under free channels. It is nonnegative because the
eigenvalues of a density operator are between zero and one, which gives
$0\le q_\alpha(\rho),q_{1-\alpha}(\rho)\le1$ and hence
$0\le\widetilde\cT_\alpha(\rho)\le1$, with the lower bound attained
at $\rho=f_1$ and the upper bound at every maximally textured state. It vanishes at the textureless state because
$f_1^\alpha=f_1^{1-\alpha}=f_1$, so
$q_\alpha(f_1)=q_{1-\alpha}(f_1)=1$ and
$\widetilde\cT_\alpha(f_1)=0$. It is monotone under QST free operations
because the data-processing inequality for the Petz
overlap~\cite{Petz1986} states that, for any CPTP map $\mathcal{E}$,
\begin{equation}
 \begin{aligned}
 \Tr[\mathcal{E}(\rho)^\alpha\mathcal{E}(\sigma)^{1-\alpha}]
 \ge\Tr(\rho^\alpha\sigma^{1-\alpha}).
 \end{aligned}
 \label{eq:petz-overlap}
\end{equation}
Setting $\sigma=f_1$ and using $\Lambda(f_1)=f_1$ gives
$q_\alpha(\Lambda(\rho))\ge q_\alpha(\rho)$ for every QST free operation.
The same inequality with $\alpha$ replaced by $1-\alpha$ gives
$q_{1-\alpha}(\Lambda(\rho))\ge q_{1-\alpha}(\rho)$.
Each factor is nonnegative and cannot decrease under the channel,
so their product cannot decrease either. Consequently,
\begin{equation}
 \widetilde\cT_\alpha(\Lambda(\rho))
 \le\widetilde\cT_\alpha(\rho).
 \label{eq:corrected-monotonicity}
\end{equation}

The WYD texture family is faithful: its only zero is the textureless state.
Let us write $p=\mel{f_1}{\rho}{f_1}$ and use the spectral decomposition
$\rho=\sum_j\lambda_j\dyad{e_j}$. The weights
$w_j=\abs{\braket{e_j}{f_1}}^2$ satisfy $\sum_jw_j=1$, so
$q_\alpha(\rho)=\sum_jw_j\lambda_j^\alpha$ and
$p=\sum_jw_j\lambda_j$. Jensen's inequality for the concave function $u^\alpha$ then gives
$q_\alpha(\rho)\le(\sum_jw_j\lambda_j)^\alpha=p^\alpha$. The same argument gives
$q_{1-\alpha}(\rho)\le p^{1-\alpha}$, and consequently
\begin{equation}
 \widetilde\cT_\alpha(\rho)
 =1-q_\alpha(\rho)q_{1-\alpha}(\rho)\ge1-p.
 \label{eq:corrected-faithfulness-bound}
\end{equation}
If the monotone vanishes, then $p=1$. Positivity and unit trace then force
$\rho=f_1$, since all population lies in this one-dimensional subspace.
The converse follows from $\widetilde\cT_\alpha(f_1)=0$. The lower bound
$1-p$ in \Eref{eq:corrected-faithfulness-bound} is the geometric texture
$\cT_g(\rho)$. The same reasoning therefore establishes faithfulness
for any texture monotone bounded below by $\cT_g$.

The WYD texture family is also convex. Lieb's concavity theorem~\cite{Lieb1973}
states that
$\rho\mapsto\Tr(\rho^\alpha f_1\rho^{1-\alpha}f_1)$ is concave.
This trace equals $q_\alpha(\rho)q_{1-\alpha}(\rho)$ because $f_1$ has
rank one. Subtracting it from unity gives, for any ensemble
$\{p_j,\rho_j\}$,
\begin{equation}
 \widetilde\cT_\alpha\Bigl(\sum_j p_j\rho_j\Bigr)
 \le\sum_j p_j\widetilde\cT_\alpha(\rho_j).
 \label{eq:corrected-convexity}
\end{equation}
Thus classical mixing, with the preparation label discarded, cannot
increase the texture above the ensemble average.

The WYD texture family can also be expressed in terms of the texture
monotones $\mathcal T^R_r$ based on the sandwiched R\'enyi relative
entropy, where $1/2\le r<1$ is their R\'enyi
parameter~\cite{Cao2026,ChenLei2026}. Defining $A_s$ through these
monotones gives
\begin{equation}
 \begin{gathered}
 A_s(\rho):=q_s(\rho)^{1/s}
 =1-\frac{s}{1+s}\mathcal T^R_{1/(1+s)}(\rho),
 \quad 0<s<1,\\
 \widetilde\cT_\alpha(\rho)
 =1-A_\alpha(\rho)^\alpha A_{1-\alpha}(\rho)^{1-\alpha}.
 \end{gathered}
 \label{eq:wyd-renyi-relation}
\end{equation}
Its nonnegativity, faithfulness, monotonicity under free channels,
and convexity can therefore also be recovered from the corresponding
properties of these monotones and the concavity of the weighted
geometric mean. In particular,
$\widetilde\cT_{1/2}=\mathcal T^R_{2/3}/3$.
This relation places the WYD texture family within
the existing framework of texture quantification. 
The comparison above does not establish strong monotonicity, which
we examine next.

At $\alpha=1/2$, the WYD texture monotone is strongly monotone: retaining
the outcomes of a free operation cannot increase its average value. It follows 
directly from the selective affinity inequality~\cite{Yu2017}. For any pair of states
$\rho,\sigma$ and any Kraus decomposition of a CPTP map, it states that
$\Tr(\sqrt\rho\sqrt\sigma)
\le\sum_j\sqrt{p_js_j}\,\Tr(\sqrt{\rho_j}\sqrt{\sigma_j})$,
where $p_j=\Tr(K_j\rho K_j^\dagger)$, $s_j=\Tr(K_j\sigma K_j^\dagger)$,
$\rho_j=K_j\rho K_j^\dagger/p_j$, and
$\sigma_j=K_j\sigma K_j^\dagger/s_j$, with zero-probability terms
contributing zero. Since $f_1$ is pure, every
free channel satisfies $K_jf_1K_j^\dagger=r_jf_1$, with
$r_j\ge0$ and $\sum_jr_j=1$. Applying the affinity inequality to
$\rho$ and $f_1$ gives 
\begin{equation}
 q_{1/2}(\rho)
 \le\sum_j\sqrt{p_jr_j}\,q_{1/2}(\rho_j).
 \label{eq:wy-selective-affinity}
\end{equation}
Squaring both sides and applying Cauchy--Schwarz, with
$\sum_jr_j=1$, then yields
\begin{equation}
 \begin{aligned}
 q_{1/2}(\rho)^2
 &\le\Bigl[\sum_j\sqrt{p_jr_j}\,q_{1/2}(\rho_j)\Bigr]^2\\
 &\le\sum_jp_jq_{1/2}(\rho_j)^2.
 \end{aligned}
 \label{eq:wy-affinity-strong-monotonicity}
\end{equation}
Using $\widetilde\cT_{1/2}(\rho)=1-q_{1/2}(\rho)^2$ and
$\sum_jp_j=1$ then gives
\begin{equation}
 \sum_jp_j\widetilde\cT_{1/2}(\rho_j)
 \le\widetilde\cT_{1/2}(\rho).
 \label{eq:strong-monotonicity}
\end{equation}
The bound in \Eref{eq:strong-monotonicity} also holds when several
Kraus outcomes are reported as a single outcome. The corresponding
conditional state is a probability-weighted mixture of the individual
conditional states, so convexity [\Eref{eq:corrected-convexity}] ensures
that this grouping cannot increase the average texture.

Strong monotonicity fails for every $\alpha\ne1/2$, already in a
three-dimensional system. Appendix~\ref{app:wyd-strong-counterexample}
constructs a family of rank-two states $\rho_\varepsilon$, with
$0<\varepsilon<1$, and a fixed two-outcome free operation, independent
of both $\alpha$ and $\varepsilon$. The average increase in the monotone
of \Eref{eq:corrected} is
\begin{equation}
 \begin{aligned}
 \Delta_\alpha(\varepsilon)
 &:=\sum_{j=0}^1p_j\widetilde\cT_\alpha(\rho_j)
      -\widetilde\cT_\alpha(\rho_\varepsilon)\\
 &=\frac{\varepsilon^2}{4(1+\varepsilon^2)}
 \left(\varepsilon^{2\alpha-1}
       +\varepsilon^{1-2\alpha}-6\right).
 \end{aligned}
 \label{eq:strong-monotonicity-asymptotic-violation}
\end{equation}
For any fixed $\alpha\ne1/2$, one of the two powers in parentheses
diverges as $\varepsilon\to0$. Thus
$\Delta_\alpha(\varepsilon)>0$ throughout an interval of sufficiently
small positive $\varepsilon$, violating \Eref{eq:strong-monotonicity}.

For independent systems, the WYD texture monotone is subadditive, with
strict inequality whenever both states are textured.
Using the product textureless state,
$(\rho\otimes\sigma)^\alpha=\rho^\alpha\otimes\sigma^\alpha$ gives
$q_\alpha(\rho\otimes\sigma)=q_\alpha(\rho)q_\alpha(\sigma)$.
Substitution into \Eref{eq:corrected} yields the exact composition law
\begin{equation}
 \begin{aligned}
 \widetilde\cT_\alpha(\rho\otimes\sigma)
 &=\widetilde\cT_\alpha(\rho)+\widetilde\cT_\alpha(\sigma)\\
 &\quad-\widetilde\cT_\alpha(\rho)\widetilde\cT_\alpha(\sigma).
 \end{aligned}
 \label{eq:corrected-composition}
\end{equation}
The product term is nonnegative and vanishes precisely when at least
one factor is textureless. Thus no member of the family is additive on
arbitrary product states, although adding a textureless ancilla leaves
the monotone unchanged.

% *}}}
\section{A Hilbert–Schmidt-based texture monotone} % {{{*
\label{sec:hs-monotone}

The failure of the squared Hilbert--Schmidt distance can be remedied
by maximizing it over QST free operations. This construction assigns to
each state the largest value of the original candidate that can be
reached without consuming texture. Let us define
\begin{equation}
 \cT_{\mathrm{HS}}^\uparrow(\rho)
 :=\max_{\Lambda\ {\rm free}}
 \cT_{\mathrm{HS}}(\Lambda(\rho)).
 \label{eq:hs-corrected}
\end{equation}
Because the identity channel is free, the quantity
satisfies
$\cT_{\mathrm{HS}}^\uparrow(\rho)\ge\cT_{\mathrm{HS}}(\rho)$.
Physically, this quantity captures the largest departure from the textureless
state that the input can produce through free operations, measured by the
squared Hilbert--Schmidt distance. Since free
operations cannot supply additional texture, the largest attainable separation
is attributed to the texture available in the state.

The optimized Hilbert--Schmidt quantity in \Eref{eq:hs-corrected} reduces
to twice the squared trace-distance texture monotone.
The trace distance between a state and the textureless state,
$\cT_{\mathrm{tr}}(\rho):=D_{\mathrm{tr}}(\rho,f_1)
=\tfrac12\norm{\rho-f_1}_1$, has been established as a texture monotone~\cite{Wang2025}.
In Appendix~\ref{app:hs-optimization}, we show that
\begin{equation}
 \cT_{\mathrm{HS}}^\uparrow(\rho)
 =\frac12\norm{\rho-f_1}_1^2
 =2\cT_{\mathrm{tr}}(\rho)^2.
 \label{eq:hs-corrected-closed}
\end{equation}
The proof bounds the squared Hilbert--Schmidt distance attainable by any
free operation and constructs a free channel that attains
$\tfrac12\norm{\rho-f_1}_1^2$.
Thus the optimized quantity gives the same ordering of resource states as
the trace-distance monotone, since it is a strictly increasing function of
that monotone.

The monotone inherits nonnegativity, faithfulness, and
monotonicity from the trace-distance monotone. The latter satisfies
$\cT_{\mathrm{tr}}(\rho)\ge0$, with equality only at $\rho=f_1$,
and the function $t\mapsto2t^2$ is strictly increasing for $t\ge0$. Hence
\begin{equation}
 \begin{gathered}
 \cT_{\mathrm{HS}}^\uparrow(\rho)\ge0,\\
 \cT_{\mathrm{HS}}^\uparrow(\rho)=0
 \quad\Longleftrightarrow\quad\rho=f_1.
 \end{gathered}
 \label{eq:hs-corrected-faithfulness}
\end{equation}
For every free channel $\Lambda$, monotonicity of
$\cT_{\mathrm{tr}}$ likewise gives
\begin{equation}
 \begin{aligned}
 \cT_{\mathrm{HS}}^\uparrow(\Lambda(\rho))
 &=2\cT_{\mathrm{tr}}(\Lambda(\rho))^2\\
 &\le2\cT_{\mathrm{tr}}(\rho)^2
 =\cT_{\mathrm{HS}}^\uparrow(\rho).
 \end{aligned}
 \label{eq:hs-corrected-monotonicity}
\end{equation}
Thus the maximization restores monotonicity independently of the contractivity
properties of the Hilbert--Schmidt norm.

The monotone also inherits convexity. For any ensemble
$\{p_j,\rho_j\}$, convexity of the trace-distance monotone and of
$t\mapsto t^2$ gives
\begin{equation}
 \begin{aligned}
 \cT_{\mathrm{HS}}^\uparrow\Bigl(\sum_jp_j\rho_j\Bigr)
 &=2\cT_{\mathrm{tr}}\Bigl(\sum_jp_j\rho_j\Bigr)^2\\
 &\le2\Bigl(\sum_jp_j\cT_{\mathrm{tr}}(\rho_j)\Bigr)^2\\
 &\le\sum_jp_j\,2\cT_{\mathrm{tr}}(\rho_j)^2\\
 &=\sum_jp_j\cT_{\mathrm{HS}}^\uparrow(\rho_j).
 \end{aligned}
 \label{eq:hs-corrected-convexity}
\end{equation}
The optimization therefore restores a faithful, convex texture monotone,
while its relation to trace distance makes its evaluation explicit.

The corrected monotone coincides with the original candidate for all qubit
states and for pure states in any dimension.
Both $\rho$ and $f_1$ have unit trace, so their difference $X=\rho-f_1$
satisfies $\Tr X=1-1=0$. For qubits, $X$ is a Hermitian
$2\times2$ matrix and therefore has eigenvalues
$a$ and $-a$, so its squared Hilbert--Schmidt norm equals half
its squared trace norm. Combining \Eref{eq:hs-qubit-trace-distance} and
\Eref{eq:hs-corrected-closed} gives
\begin{equation}
 \cT_{\mathrm{HS}}^\uparrow(\rho)
 =\cT_{\mathrm{HS}}(\rho),\qquad D=2.
 \label{eq:hs-corrected-qubit}
\end{equation}
The same equality holds for pure states in any dimension, because the difference
of two rank-one projectors has at most one positive and one negative eigenvalue.
The two quantities can therefore differ only for mixed states in dimensions $D\ge3$.

Unlike the WYD texture monotone at $\alpha=1/2$, the
Hilbert--Schmidt monotone does not satisfy strong monotonicity.
Let us construct a qubit counterexample. We take a normalized
$\ket g\perp\ket{f_1}$ and consider $\rho=pf_1+(1-p)\dyad g$, where $0<p<1$.
Let us apply the QST free operation in \Eref{eq:measure-prepare-free} with
$\sigma=\dyad g$ and retain the outcome of its defining measurement.
The two outcomes have probabilities $p$ and $1-p$ and prepare
$f_1$ and $\dyad g$, respectively. Since
$\cT_{\mathrm{HS}}^\uparrow(\rho)=2(1-p)^2$,
$\cT_{\mathrm{HS}}^\uparrow(f_1)=0$, and
$\cT_{\mathrm{HS}}^\uparrow(\dyad g)=2$, the average
postmeasurement texture satisfies
\begin{equation}
 \begin{aligned}
 \sum_{j=0}^1p_j
 \cT_{\mathrm{HS}}^\uparrow(\rho_j)
 &=2(1-p)\\
 &>2(1-p)^2\\
 &=\cT_{\mathrm{HS}}^\uparrow(\rho).
 \end{aligned}
 \label{eq:hs-corrected-strong-violation}
\end{equation}
Thus deterministic free operations cannot increase the monotone,
but conditioning on their outcomes can increase its average value.
This failure occurs already in the qubit regime, where the proposed
monotone and the original Hilbert--Schmidt candidate coincide.

For composite systems, the monotone is invariant under adding a
textureless ancilla. With the same notation as in the preceding section,
let us write the joint textureless state as
$f_1=f_1^{(A)}\otimes f_1^{(B)}$. Multiplicativity of the trace norm
under tensor products gives
\begin{equation}
 \begin{aligned}
 \cT_{\mathrm{HS}}^\uparrow(\rho_A\otimes f_1^{(B)})
 &=\frac12
   \norm{(\rho_A-f_1^{(A)})\otimes f_1^{(B)}}_1^2\\
 &=\cT_{\mathrm{HS}}^\uparrow(\rho_A).
 \end{aligned}
 \label{eq:hs-corrected-ancilla}
\end{equation}
Thus introducing a free ancilla does not change the texture
assigned to the original state.
This invariance allows texture to be compared before and after embedding a
system into a larger register initialized in a textureless state.

The monotone is not additive on general product states and
satisfies neither universal subadditivity nor universal
superadditivity. For example, two pure states, each with
overlap $1/2$ with its local textureless state, have individual
textures equal to $1$, whereas their product has
texture $3/2$, strictly below their sum. Conversely, for two
states of the form
$\rho_X=(f_1^{(X)}+\dyad{g_X})/2$, with
$\ket{g_X}\perp\ket{f_1^{(X)}}$, each local texture equals $1/2$,
while their product has texture $9/8$, strictly above their
sum. The tensor-product behavior therefore depends on the
states beyond their individual texture values. 

% *}}}
\section{Texture bounds at fixed overlap with the textureless state} % {{{*
\label{sec:texture-bounds}

Depending on the texture monotone, states with the same overlap with
the textureless state can be assigned different values.
Zero overlap gives the global maximum of every texture monotone by
\Eref{eq:orthogonal-maximality}, whereas unit overlap admits only the
textureless state. At intermediate overlaps, the WYD family retains
spectral information beyond this population [\Eref{eq:corrected}], and
the corrected Hilbert--Schmidt monotone depends on the trace distance
from the textureless state [\Eref{eq:hs-corrected-closed}].
We therefore ask for the exact range of each monotone at fixed overlap.

The block decomposition relative to $\ket{f_1}$ separates the fixed
population $p$ from the remaining state variables. Choose an orthonormal
basis $\{\ket{f_1},\ket{g_1},\ldots,\ket{g_{D-1}}\}$, where
$\{\ket{g_j}\}_{j=1}^{D-1}$ spans the orthogonal complement
$\{\ket{f_1}\}^{\perp}$. In this basis, the state reads
\begin{equation}
 \rho=\vcenter{\hbox{%
  \includegraphics[width=0.38\columnwidth]
   {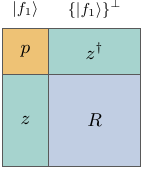}%
 }},
 \label{eq:texture-block-decomposition}
\end{equation}
where $p=\mel{f_1}{\rho}{f_1}=\Sigma(\rho)/D$
[\Eref{eq:prelim-textureless}], $z$ contains the coherences with the
orthogonal complement, and $R\succeq0$ is the block supported on that
subspace, with $\Tr R=1-p$. Colors distinguish these blocks and do not
encode numerical entries.
Maximally textured states occupy only the orthogonal block in
\Eref{eq:texture-block-decomposition}. Indeed,
$\rho\in\mathcal S_\perp$ is equivalent to $p=0$
[\Sref{sec:reachability}], and positivity then forces $z=0$.

Averaging over the unitaries that preserve $f_1$ selects a unique
symmetric representative at each overlap $p$ (Appendix~\ref{app:twirl}).
These unitaries have the form $U=e^{i\phi}f_1+V_Q$, with
$Q=\mathbb I-f_1$ and $V_Q$ unitary on its support, and form the
isotropy group $G_{f_1}\cong\mathrm U(1)\times\mathrm U(D-1)$.
Every texture monotone is invariant under them, since their inverses
are also free [\Eref{eq:prelim-monotone}]. Their Haar average defines the
twirl~\cite{VollbrechtWerner2001,Watrous2018}
\begin{equation}
 \mathcal W(\rho):=\int_{G_{f_1}}\dd U\,U\rho U^\dagger,
 \label{eq:twirl}
\end{equation}
where $\dd U$ is the normalized Haar measure. The twirl itself is a free operation because
each unitary fixes $f_1$, and it preserves $p$. Invariance under the
group sets $z=0$ and makes $R$ proportional to the identity on the
orthogonal complement, giving
\begin{equation}
 \mathcal W(\rho)=\rho_p:=pf_1+(1-p)\frac{Q}{D-1}.
 \label{eq:isotropic-axis}
\end{equation}
Appendix~\ref{app:twirl} establishes these properties and describes
the twirl's action on the remaining operator components.

\begin{figure}[t]
 \centering
 \includegraphics[width=\columnwidth]{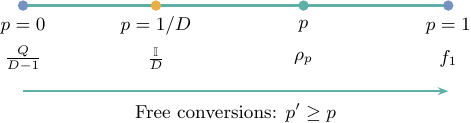}
 \caption{The isotropic axis for a system of dimension $D\ge2$.
 The segment joins the maximally textured state $Q/(D-1)$ to the
 textureless state $f_1$ and contains $\mathbb I/D$ at $p=1/D$.
 Each point is the state $\rho_p$ of \Eref{eq:isotropic-axis},
 with overlap $p=\Tr(\rho_p f_1)$.
 Free conversions along the axis increase $p$, so every texture
 monotone is nonincreasing from left to right.
 The left endpoint $Q/(D-1)$ has rank $D-1$ and is pure only for $D=2$.
 Positions are schematic and not drawn to scale.}
 \label{fig:isotropic-axis}
\end{figure}

The symmetric representatives $\rho_p$ in \Eref{eq:isotropic-axis} form an isotropic axis connecting
the maximally textured set to the unique textureless state.
The straight segment in \Fref{fig:isotropic-axis} joins $Q/(D-1)$ at
$p=0$ to $f_1$ at $p=1$ through $\rho_{1/D}=\mathbb I/D$
[\Eref{eq:isotropic-axis}]. Its orthogonal endpoint is maximally
textured by \Sref{sec:reachability}, although mixed for $D>2$;
$\mathbb I/D$ is an interior textured state. The axis is fixed by
$f_1$ and $D$, without choosing a texture monotone.

The free twirl makes $\rho_p$ a minimizer of every texture monotone
among states with the same overlap $p$. Indeed, \Eref{eq:isotropic-axis}
and monotonicity give
\begin{equation}
 \cT(\rho)\ge\cT(\mathcal W(\rho))=\cT(\rho_p).
 \label{eq:twirl-texture-lower-bound}
\end{equation}

Every state commuting with $f_1$ also attains this minimum at its
overlap $p$.
In \Eref{eq:texture-block-decomposition}, commutation requires only
$z=0$, whereas membership in the axis additionally requires
$R=(1-p)Q/(D-1)$. These conditions coincide for qubits but differ for
$D>2$. Nevertheless, the channel in \Eref{eq:measure-prepare-free}
with an orthogonal target interconverts every commuting state with
$\rho_p$ at fixed $p$, so all texture monotones assign them the same
value. Thus the minimizing representative is unique under the symmetry
requirement, but need not be the only state attaining the minimum.

For the WYD family, the minimum selected by the twirl gives the lower
bound $1-p$ [\Eref{eq:corrected-faithfulness-bound}], while the spectral
moments determine a tight upper bound. To obtain the upper bound,
let us write
$\rho=\sum_j\lambda_j\dyad{e_j}$, with weights
$w_j=\abs{\braket{e_j}{f_1}}^2$. Then
$p=\sum_jw_j\lambda_j$ and
$q_s(\rho)=\sum_jw_j\lambda_j^s$.
For $0<s<1$, $\lambda_j^s\ge\lambda_j$ gives $q_s(\rho)\ge p$.
Combining this with the moment upper bound used in the proof of
\Eref{eq:corrected-faithfulness-bound} gives
\begin{equation}
 p\le q_s(\rho)\le p^s,\qquad 0<s<1.
 \label{eq:bounds-moments}
\end{equation}
Applying \Eref{eq:bounds-moments} at $\alpha$ and $1-\alpha$
gives $p^2\le q_\alpha(\rho)q_{1-\alpha}(\rho)\le p$, and hence
\begin{equation}
 1-p\le\widetilde\cT_\alpha(\rho)\le1-p^2.
 \label{eq:bounds-corrected}
\end{equation}
The bounds are independent of $\alpha$, although individual
mixed-state values can depend on it.

\begin{figure*}[t]
 \centering
 \includegraphics[width=\textwidth]
 {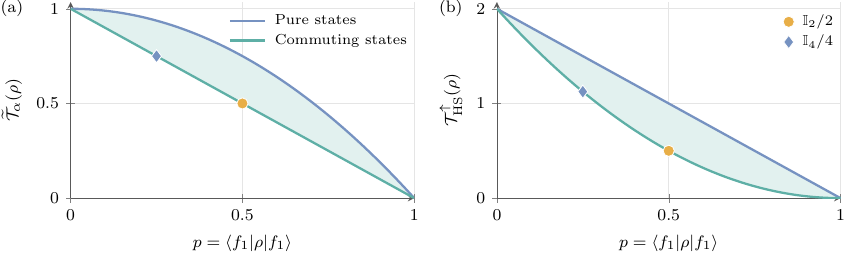}
 \caption{Tight overlap bounds for (a) the WYD texture monotones,
 \Eref{eq:bounds-corrected}, and (b) the corrected Hilbert--Schmidt
 monotone, \Eref{eq:bounds-hs-corrected}.
 The WYD bounds hold for every $0<\alpha<1$.
 Shaded regions give all attainable values
 at fixed $p=\mel{f_1}{\rho}{f_1}$ in every dimension $D\ge2$,
 as established in \Sref{sec:texture-bounds} together with the
 saturation conditions: for $0<p<1$, upper bounds are attained
 exactly by pure states, and lower bounds exactly by states
 commuting with $f_1$.
 Circles and diamonds indicate $\mathbb I_2/2$ and $\mathbb I_4/4$,
 respectively.}
 \label{fig:texture-bounds}
\end{figure*}

The corrected Hilbert--Schmidt bounds follow from its trace-distance
representation in \Eref{eq:hs-corrected-closed}. The variational
characterization in \Eref{eq:hs-corrected-eigenvalue} and the eigenvalue
argument leading to \Eref{eq:bounds-hs-equality} below give
\begin{equation}
 1-p\le\cT_{\mathrm{tr}}(\rho)\le\sqrt{1-p}.
\end{equation}
Squaring and using \Eref{eq:hs-corrected-closed} yields
\begin{equation}
 2(1-p)^2\le\cT_{\mathrm{HS}}^\uparrow(\rho)\le2(1-p).
 \label{eq:bounds-hs-corrected}
\end{equation}
\Fref{fig:texture-bounds} summarizes
\Eref{eq:bounds-corrected} and \Eref{eq:bounds-hs-corrected}.
The left and right panels plot $\widetilde\cT_\alpha(\rho)$ and
$\cT_{\mathrm{HS}}^\uparrow(\rho)$ against $p$, respectively.
The blue upper curves are $1-p^2$ and
$2(1-p)$, while the teal lower curves are $1-p$ and $2(1-p)^2$,
respectively. The shaded vertical intervals represent all attainable
values between these boundaries. The single WYD panel applies to every $0<\alpha<1$:
changing $\alpha$ can change a mixed state's value within the shaded
region, but leaves the boundaries unchanged. To establish the exact
attainable ranges, we now identify the states that reach each curve
and show that all intermediate values are accessible.

Both lower bounds in \Eref{eq:bounds-corrected} and \Eref{eq:bounds-hs-corrected}
are attained by every state with no coherences between
the textureless state and its orthogonal complement. Let us take
\begin{equation}
 \rho_{\mathrm{low}}=pf_1+(1-p)\rho_\perp,
 \label{eq:bounds-lower-state}
\end{equation}
with $\rho_\perp$ any maximally textured state. Orthogonality of the supports gives
$q_s(\rho_{\mathrm{low}})=p^s$ and
$\rho_{\mathrm{low}}-f_1=(p-1)f_1+(1-p)\rho_\perp$.
Thus $\cT_{\mathrm{tr}}(\rho_{\mathrm{low}})=1-p$.
Strict concavity of the scalar function
$\lambda\mapsto\lambda^\alpha$ makes this condition necessary as well:
equality in Jensen's inequality requires every
eigenvalue with $w_j>0$ to equal $p$. Equivalently,
$\rho\ket{f_1}=p\ket{f_1}$, or $\comm{\rho}{f_1}=0$.
For the Hilbert--Schmidt lower bound, the same condition follows from
the variational characterization of the negative eigenvalue of
$\rho-f_1$: equality in $-\lambda_{\min}(\rho-f_1)\ge1-p$
requires $\ket{f_1}$ to be its eigenvector.
Thus the commuting states identified through the twirl
[\Eref{eq:twirl-texture-lower-bound}] are precisely the minimizers
of both monotones.
Commutation with $f_1$ does not require diagonality in the
computational basis. For one qubit, the minimizing state is
$p\dyad{+}+(1-p)\dyad{-}$; for two qubits, $\rho_\perp$ can be any
density operator on the three-dimensional complement of $\ket{++}$.

Both upper bounds in \Eref{eq:bounds-corrected} and \Eref{eq:bounds-hs-corrected} are
attained by every pure state with the prescribed overlap. These states can be written as
$\rho_{\mathrm{up}}=\dyad{\psi_p}$, with
$\ket{\psi_p}=\sqrt p\ket{f_1}+\sqrt{1-p}\ket{g}$ for any normalized
$\ket{g}\perp\ket{f_1}$.
Since $\rho_{\mathrm{up}}$ is a rank-one projector,
$\rho_{\mathrm{up}}^s=\rho_{\mathrm{up}}$ for every $s>0$.
Thus the definition in \Eref{eq:corrected} gives
$q_\alpha(\rho_{\mathrm{up}})=p$, while
$\rho_{\mathrm{up}}-f_1$ has eigenvalues
$\pm\sqrt{1-p}$ and otherwise zeros. The two monotones therefore equal
$1-p^2$ and $2(1-p)$, respectively.

For the WYD family, saturation requires
$q_\alpha q_{1-\alpha}=p^2$. Since each factor is at least $p$,
both must equal $p$. Equality in
$q_\alpha-p=\sum_jw_j(\lambda_j^\alpha-\lambda_j)\ge0$
then requires $\lambda_j\in\{0,1\}$ whenever $w_j>0$.
The Hilbert--Schmidt bound has the same equality condition.
Indeed, writing $a=\cT_{\mathrm{tr}}(\rho)=-\lambda_{\min}(\rho-f_1)>0$, its eigenvalue
equation and the scalar inequality
$1/(a+\lambda)\le(1-\lambda)/a+\lambda/(a+1)$ give
\begin{equation}
 1=\sum_j\frac{w_j}{a+\lambda_j}
 \le\frac{1-p}{a}+\frac{p}{a+1}.
 \label{eq:bounds-hs-equality}
\end{equation}
This is equivalent to $a^2\le1-p$, with equality only when
$\lambda_j\in\{0,1\}$ for every $w_j>0$.
For either monotone, $p>0$ thus requires an eigenvalue equal to one;
unit trace then forces $\rho$ to be pure.

For maximally textured states ($p=0$), the lower and
upper curves in each panel of \Fref{fig:texture-bounds} meet, giving $\widetilde\cT_\alpha=1$ and $\cT_{\mathrm{HS}}^\uparrow=2$.
The free channel in \Eref{eq:measure-prepare-free} can transform any such
state into any target state, consistent with the maximal texture values above.
For $D>2$, maximally textured states can also be mixed.
At $p=1$, the only state is $f_1$, and both curves meet at zero.

For a system of arbitrary dimension $D\ge2$, every value between the
bounds can be realized within a two-dimensional subspace of its full
Hilbert space. Let us choose a normalized $\ket{g}\perp\ket{f_1}$ and an
orthonormal basis beginning with $\{\ket{f_1},\ket{g}\}$. In this basis,
consider, for $0\le t\le1$,
\begin{equation}
 \rho_{p,t}=\begin{pmatrix}
 p&t\sqrt{p(1-p)}\\
 t\sqrt{p(1-p)}&1-p
 \end{pmatrix}\oplus 0_{D-2},
 \label{eq:bounds-interpolating-state}
\end{equation}
where $0_{D-2}$ is the zero operator on the remaining orthogonal
subspace, omitted when $D=2$. Thus the $2\times2$ block specifies the
occupied subspace, while the full state acts on the $D$-dimensional
system. With $\rho_\perp=\dyad{g}$ in \Eref{eq:bounds-lower-state},
this family equals the convex combination
$\rho_{p,t}=(1-t)\rho_{\mathrm{low}}+t\rho_{\mathrm{up}}$.
It preserves the overlap and attains both lower bounds at $t=0$
and both upper bounds at $t=1$. Along this family,
$\cT_{\mathrm{HS}}^\uparrow(\rho_{p,t})
=2[(1-p)^2+t^2p(1-p)]$.
Continuity then guarantees every
intermediate value for each monotone separately. Consequently, the
shaded regions in \Fref{fig:texture-bounds} are exact for every
$D\ge2$. 

As a reference, the maximally mixed state $\rho_*=\mathbb I/D$ has
$p_*=1/D$ and $q_\alpha(\rho_*)=D^{-\alpha}$, giving
$\widetilde\cT_\alpha(\rho_*)=1-1/D$ and
$\cT_{\mathrm{HS}}^\uparrow(\rho_*)=2(1-1/D)^2$.
This state attains the lower bounds in both panels and is represented in
\Fref{fig:texture-bounds} by gold circles for $D=2$ and blue
diamonds for $D=4$. The circles lie at
$(p,\cT)=(1/2,1/2)$ for both monotones, while the diamonds lie at
$(1/4,3/4)$ in the WYD panel and $(1/4,9/8)$ in the
Hilbert--Schmidt panel.

% *}}}
\section{Conclusions} % {{{*
\label{sec:conclusions}
We use reachability under QST free operations to constrain the
quantification of texture. We call states supported on the orthogonal
complement of the textureless state maximally textured. Every maximally
textured state can reach any target state through a free operation, so
every texture monotone must assign the same global maximum to all such
states. In particular, any nonnegative texture monotone that vanishes
on even one maximally textured state must vanish on every state.
We use this requirement to show that a proposed quantifier based on the
WYD skew information of a state with respect to the textureless state
is not a texture monotone: it vanishes on maximally textured states
but is positive on some states reachable from them by free operations.
A proposed quantifier given
by the squared Hilbert--Schmidt distance from the textureless state
also fails monotonicity in dimensions $D\ge3$: a free operation can
increase this distance by purifying a maximally textured state.
Both failures reflect the requirement that all states orthogonal to the
textureless state carry the same maximal texture, which in dimensions
greater than two also applies to mixed states because the orthogonal
subspace has dimension greater than one.

We resolve these monotonicity failures by constructing valid texture
monotones from both proposed quantifiers.
Adding the population outside the textureless state
to the skew information yields a faithful, convex family of texture
monotones. Maximizing the proposed
Hilbert--Schmidt candidate over free operations
also gives a faithful, convex monotone and identifies twice the squared
trace-distance texture as the largest squared Hilbert--Schmidt distance
attainable through free processing.

Finally, we determine the exact ranges of both corrected monotones
at fixed overlap with the textureless state.
The free twirl, obtained by averaging over free unitaries, preserves this
overlap, removes coherences between the textureless state and its orthogonal complement, and
distributes the remaining population uniformly within that complement.
The states with these properties define the isotropic axis.
The resulting states attain both lower bounds. More generally, the lower bounds are
attained precisely by states commuting with the textureless state,
whereas at interior overlaps the upper bounds require pure states.
Within the subspace spanned by the textureless state and one orthogonal
state, continuously increasing the coherence at fixed populations
connects a lower-bound state to a pure upper-bound state.
This construction realizes every intermediate value for each monotone
separately.

% *}}}
\section*{Acknowledgements} % {{{*
J.A.d.L. and M.G. acknowledge SECIHTI through funding for graduate studies.
J.A.d.L. acknowledges support by UNAM-PAPIIT IG101324 and SECIHTI CBF-2025-I-1548.
The authors acknowledge the use of OpenAI's ChatGPT (GPT-6.1) and Codex as assistive tools to explore possible approaches to the problems addressed in this work, identify relevant literature and assess the novelty of the results, and cross-check mathematical derivations. All AI-assisted suggestions were critically evaluated by the authors, who independently verified the relevant references, derivations, and scientific conclusions. The authors take full responsibility for the content and results of this work.
% *}}}

\appendix
\section{Optimal free channel for the Hilbert--Schmidt texture monotone} % {{{*
\label{app:hs-optimization}

The variational definition in \Eref{eq:hs-corrected} asks how large
the squared Hilbert--Schmidt distance from $f_1$ can become under a
free channel. We first bound this distance by twice the squared
trace distance of the input. We then show that the bound is attainable:
the spectral structure of $\rho-f_1$ allows a free channel to
concentrate its positive and negative parts into a two-dimensional
subspace. This construction proves \Eref{eq:hs-corrected-closed}
and also yields an expression for the monotone in terms of a single
eigenvalue.

To prove \Eref{eq:hs-corrected-closed}, let $X=\rho-f_1$. Every free channel
satisfies $\Lambda(X)=\Lambda(\rho)-f_1$. For any traceless
Hermitian operator $Y$, its positive and negative parts have equal
trace, which implies
\begin{equation}
 \norm{Y}_2^2\le\frac12\norm{Y}_1^2.
 \label{eq:hs-traceless-bound}
\end{equation}
Applying this inequality to $Y=\Lambda(X)$ and using
contractivity of the trace norm under CPTP maps gives
\begin{equation}
 \begin{aligned}
 \norm{\Lambda(X)}_2^2
 &\le\frac12\norm{\Lambda(X)}_1^2\\
 &\le\frac12\norm{X}_1^2.
 \end{aligned}
 \label{eq:hs-corrected-upper-bound}
\end{equation}
It remains to construct a free channel that saturates this bound.

The structure of $X$ permits such a construction. For every vector
orthogonal to $\ket{f_1}$, the quadratic form of $X$ is nonnegative
because it coincides with that of $\rho$. Consequently, $X$ has
at most one negative eigenvalue. Since $\Tr X=0$, for $\rho\ne f_1$
there is exactly one~\cite{LiuWangYu2026}, and we may write
\begin{equation}
 X=X_+-a\dyad{n},\qquad
 \Tr X_+=a=\frac12\norm{X}_1,
 \label{eq:hs-jordan-decomposition}
\end{equation}
where $X_+\succeq0$ and its support is orthogonal to $\ket n$.

Let $\Pi_+$ project onto the support of $X_+$, and choose
$\ket u=\Pi_+\ket{f_1}/\norm{\Pi_+\ket{f_1}}$ whenever the
numerator is nonzero. Otherwise, choose any normalized vector
in that support. Positivity of $\rho=X+f_1$ implies that
$\ket{f_1}$ has no component in $\ker X$: for any normalized
$\ket z\in\ker X$, the principal minor of $\rho$ on
$\operatorname{span}\{\ket n,\ket z\}$ has determinant
$-a|\braket{z}{f_1}|^2$, which must be nonnegative. Therefore,
$\braket{z}{f_1}=0$, and $\ket{f_1}$ belongs to
$\mathcal S=\operatorname{span}\{\ket n,\ket u\}$.

This observation allows us to concentrate both spectral parts
of $X$ into a two-dimensional subspace without disturbing the
textureless state. Let $P=\dyad n+\dyad u$ and let
$\{\ket{e_j}\}_{j=1}^{D-2}$ be an orthonormal basis of
$\mathcal S^\perp$. Consider the channel with Kraus operators
\begin{equation}
 K_0=P,\qquad
 K_j=\ket u\bra{e_j},\quad j=1,\ldots,D-2.
 \label{eq:hs-optimal-kraus}
\end{equation}
They satisfy $\sum_jK_j^\dagger K_j=\mathbb I$, so the map is
CPTP. Furthermore, $K_0\ket{f_1}=\ket{f_1}$ and
$K_j\ket{f_1}=0$ for $j\ge1$, showing that the channel is free.

Denoting this channel by $\Lambda_*$, its action on $X$ is
\begin{equation}
 \Lambda_*(X)
 =a(\dyad u-\dyad n).
 \label{eq:hs-optimal-output}
\end{equation}
The output has two nonzero eigenvalues, $a$ and $-a$, and hence
\begin{equation}
 \begin{aligned}
 \norm{\Lambda_*(X)}_2^2
 &=2a^2\\
 &=\frac12\norm{\rho-f_1}_1^2.
 \end{aligned}
\end{equation}
The upper bound in \Eref{eq:hs-corrected-upper-bound} is
therefore attainable. For $\rho=f_1$, both sides vanish,
completing the proof of \Eref{eq:hs-corrected-closed}.

For a pure reference state, the trace distance satisfies
$\cT_{\mathrm{tr}}(\rho)=-\lambda_{\min}(\rho-f_1)$~\cite{LiuWangYu2026}.
Combining this identity with \Eref{eq:hs-corrected-closed} gives
\begin{equation}
 \cT_{\mathrm{HS}}^\uparrow(\rho)
 =2\lambda_{\min}(\rho-f_1)^2.
 \label{eq:hs-corrected-eigenvalue}
\end{equation}
Here $\lambda_{\min}(\rho-f_1)\le0$, with equality only for
$\rho=f_1$. Thus evaluating the monotone requires
neither an optimization over channels nor the full spectrum
of $\rho-f_1$.

% *}}}
\section{The free twirl} % {{{*
\label{app:twirl}

This appendix collects the properties of the twirl in \Eref{eq:twirl}
used in \Sref{sec:texture-bounds}. Let $G$ be a compact group of
unitaries with normalized Haar measure $\dd U$, and
$\mathcal W_G(\rho)=\int_G\dd U\,U\rho U^\dagger$~\cite{Watrous2018}. The
map $\mathcal W_G$ is CPTP, being a convex mixture of unitary channels.
Invariance of the Haar measure gives $V\mathcal W_G(X)V^\dagger=\mathcal W_G(X)$
for every $V\in G$ and every operator $X$, while $\mathcal W_G(X)=X$
whenever $X$ commutes with $G$; hence $\mathcal W_G\circ\mathcal W_G=\mathcal W_G$.
As an idempotent linear map, $\mathcal W_G$ has only $1$ and $0$
as possible eigenvalues;
its eigenvalue-$1$ subspace consists of the operators commuting with $G$,
and its kernel consists of the operators erased by the average.

For $G=G_{f_1}$, an operator $A$ commuting with $e^{i\phi}f_1+Q$ for every
$\phi$ satisfies $f_1AQ=QAf_1=0$. Its block $QAQ$ commutes with every
unitary on the orthogonal complement and is therefore proportional to $Q$
by Schur's lemma. The eigenvalue-$1$ subspace is thus
$\operatorname{span}\{f_1,Q\}$. Since $f_1$ and $Q$ commute with
$G_{f_1}$, the twirl preserves $\Tr(\rho f_1)$ and $\Tr(\rho Q)$, which
gives \Eref{eq:isotropic-axis}. The eigenvalue-$0$ subspace contains all
operators $X$ with $\Tr(Xf_1)=\Tr(XQ)=0$: in the notation of
\Eref{eq:texture-block-decomposition}, the coherences $z$ and the
traceless part of the orthogonal block. Every state therefore splits as
$\rho=\mathcal W(\rho)+[\rho-\mathcal W(\rho)]$, where the first term keeps
the overlap $p$ and the second contains everything the twirl erases.

The twirl is free, since each $U\in G_{f_1}$ fixes $f_1$, and it is
unital. The same channel admits two further
realizations. First, \Eref{eq:isotropic-axis} coincides with the
measure-and-prepare map of \Eref{eq:measure-prepare-free} for
$\sigma=Q/(D-1)$. Second, it equals the average over the $D!$ permutations
of an orthonormal basis $\{\ket j\}$ in which
$\ket{f_1}=D^{-1/2}\sum_j\ket j$: a matrix commuting with all permutations
has constant diagonal and constant off-diagonal entries and therefore lies
in $\operatorname{span}\{\mathbb I,f_1\}$. The analogous construction in
entanglement theory, twirling over $U\otimes U^*$, produces the isotropic
states~\cite{HorodeckiHorodecki1999}, which motivates the name used here.
% *}}}
\section{Failure of strong monotonicity for the WYD family} % {{{*
\label{app:wyd-strong-counterexample}

We construct the counterexample in
\Eref{eq:strong-monotonicity-asymptotic-violation} using a fixed free
operation on a three-dimensional system. Choose normalized vectors
$\ket g$ and $\ket h$ such that
$\{\ket{f_1},\ket g,\ket h\}$ is an orthonormal basis, and set
\begin{equation}
 \ket{e_0}=\frac{\ket{f_1}+\ket h}{\sqrt2},\qquad
 \ket u=\frac{\ket{f_1}-\ket h}{\sqrt2}.
 \label{eq:strong-counterexample-basis}
\end{equation}
The two Kraus operators are
\begin{equation}
 K_0=\dyad{f_1}{e_0},\qquad
 K_1=\dyad{f_1}{u}+\dyad g.
 \label{eq:strong-counterexample-kraus}
\end{equation}
Indeed,
$K_0^\dagger K_0+K_1^\dagger K_1 = \dyad{e_0} + \dyad u+\dyad g = \mathbb{I}$, 
so the channel is trace-preserving.
Moreover, $K_0\ket{f_1}=K_1\ket{f_1}=\ket{f_1}/\sqrt2$, so the channel fixes 
$f_1$.  
It distinguishes $\ket{e_0}$ from its orthogonal complement, prepares
$f_1$ for the first outcome, and maps $\ket u$ and $\ket g$ to
$\ket{f_1}$ and $\ket g$, respectively, for the second outcome.

For any $0<\varepsilon<1$, define
\begin{equation}
 \begin{aligned}
 \ket{e_\varepsilon}
 &=\sqrt\varepsilon\ket u+\sqrt{1-\varepsilon}\ket g,\\
 \rho_\varepsilon
 &=\frac{\varepsilon^2\dyad{e_0}
             +\dyad{e_\varepsilon}}{1+\varepsilon^2}.
 \end{aligned}
 \label{eq:strong-monotonicity-alternative-states}
\end{equation}
Since $\ket{e_0}$ is orthogonal to both $\ket u$ and $\ket g$, it is
orthogonal to the normalized vector $\ket{e_\varepsilon}$.
Thus \Eref{eq:strong-monotonicity-alternative-states} is a spectral
decomposition of the rank-two density operator $\rho_\varepsilon$,
with nonzero eigenvalues $\varepsilon^2/(1+\varepsilon^2)$ and
$1/(1+\varepsilon^2)$. The corresponding eigenvectors $\ket{e_0}$ and
$\ket{e_\varepsilon}$ have squared overlaps
$|\braket{f_1}{e_0}|^2=1/2$ and
$|\braket{f_1}{e_\varepsilon}|^2=\varepsilon/2$,
respectively.
The third eigenvalue is zero and contributes nothing to the spectral
moments for $s>0$. Hence the definition in \Eref{eq:corrected} gives
\begin{equation}
 q_s(\rho_\varepsilon)
 =\frac{\varepsilon^{2s}+\varepsilon}
        {2(1+\varepsilon^2)^s},\qquad 0<s<1.
 \label{eq:strong-counterexample-moments}
\end{equation}

Applying the Kraus operators in \Eref{eq:strong-counterexample-kraus}
to the input in \Eref{eq:strong-monotonicity-alternative-states}, we
calculate each outcome probability and conditional state from
\begin{equation}
 p_j=\Tr(K_j\rho_\varepsilon K_j^\dagger),\qquad
 \rho_j=\frac{K_j\rho_\varepsilon K_j^\dagger}{p_j},
 \quad j=0,1.
 \label{eq:strong-counterexample-outcome-rule}
\end{equation}
These are the Kraus outcome formulas introduced in
\Sref{sec:preliminaries}. They give
\begin{equation}
 \begin{aligned}
 p_0&=\frac{\varepsilon^2}{1+\varepsilon^2},
 &\rho_0&=f_1,\\
 p_1&=\frac{1}{1+\varepsilon^2},
 &\rho_1&=\dyad{\psi_\varepsilon},\\
 \ket{\psi_\varepsilon}
 &=\sqrt\varepsilon\ket{f_1}+\sqrt{1-\varepsilon}\ket g.
 \end{aligned}
 \label{eq:strong-counterexample-outcomes}
\end{equation}
For a pure state, \Eref{eq:corrected} gives
$\widetilde\cT_\alpha(\dyad\psi)=1-|\braket{f_1}{\psi}|^4$.
Thus $\widetilde\cT_\alpha(\rho_0)=0$ and
$\widetilde\cT_\alpha(\rho_1)=1-\varepsilon^2$, independently of
$\alpha$, and their average is
$(1-\varepsilon^2)/(1+\varepsilon^2)$.
Combining this average with \Eref{eq:strong-counterexample-moments}
yields \Eref{eq:strong-monotonicity-asymptotic-violation} exactly.

For fixed $\alpha\ne1/2$, let $\delta=|2\alpha-1|>0$.
The sign of the average increase is the sign of
$\varepsilon^{-\delta}+\varepsilon^\delta-6$, which tends to
$+\infty$ as $\varepsilon\to0$. The violation therefore holds for
all sufficiently small positive $\varepsilon$.
At $\alpha=1/2$, instead,
$\Delta_{1/2}(\varepsilon)=-\varepsilon^2/(1+\varepsilon^2)<0$,
consistent with \Eref{eq:strong-monotonicity}.
% *}}}

\bibliography{references}

\end{document}